\documentclass[aps,showpacs,amsmath,amssymb,twocolumn,nofootinbib]{revtex4-1}

\usepackage{subfigure}
\usepackage{graphicx}% Include figure files
\usepackage{dcolumn}% Align table columns on decimal point
\usepackage{bm}% bold math
\usepackage{color}
\usepackage[colorlinks=true,pdfstartview=FitV,linkcolor=blue,citecolor=blue,urlcolor=blue]{hyperref}
\usepackage[mathlines]{lineno}% Enable numbering of text and display math
\usepackage[dvipsnames]{xcolor}
\usepackage{amsmath}
\everymath{\displaystyle}
\begin{document}

\newcommand{\gc}[1]{\textcolor{red}{#1}}
\newcommand{\ad}[1]{\textcolor{blue}{#1}}
\newcommand{\add}[1]{\textcolor{Green}{#1}}

\newcommand{\beq}{\begin{equation}}
\newcommand{\eeq}{\end{equation}}
\newcommand{\beqs}{\begin{eqnarray}}
\newcommand{\eeqs}{\end{eqnarray}}

\title{Feebly Interacting $U(1)_{\rm B-L}$ Gauge Boson Warm Dark Matter and XENON1T Anomaly}

\author{Gongjun Choi,$^{1}$}
\thanks{{\color{blue}gongjun.choi@gmail.com}}

\author{Tsutomu T. Yanagida,$^{1,2}$}
\thanks{{\color{blue}tsutomu.tyanagida@ipmu.jp}}

\author{Norimi Yokozaki,$^{3}$}
\thanks{{\color{blue}n.yokozaki@gmail.com}}

\affiliation{$^{1}$ Tsung-Dao Lee 
Institute, Shanghai Jiao Tong University, Shanghai 200240, China}

\affiliation{$^{2}$ Kavli IPMU (WPI), UTIAS, The University of Tokyo,
5-1-5 Kashiwanoha, Kashiwa, Chiba 277-8583, Japan}

\affiliation{$^{3}$ Theory Center, IPNS, KEK, 1-1 Oho, Tsukuba, Ibaraki 305-0801, Japan}
\date{\today}

\begin{abstract}
The recent observation of an excess in the electronic recoil data by the XENON1T detector has drawn many attentions as a potential hint for an extension of the Standard Model (SM). Absorption of a vector boson with the mass of $m_{A'}\!\in\!(2\,{\rm keV},\!3\,{\rm keV})$ is one of the feasible explanations to the excess. In the case where the vector boson explains the dark matter (DM) population today, it is highly probable that the vector boson belongs to a class of the warm dark matter (WDM) due to its suspected mass regime. In such a scenario, providing a good fit for the excess, the kinetic mixing $\kappa\!\sim\!10^{-15}$ asks for a non-thermal origin of the vector DM. In this letter, we consider a scenario where the gauge boson is nothing but the $U(1)_{\rm B-L}$ gauge boson and its non-thermal origin is attributed to the decay of the coherently oscillating scalar of which condensation induces the spontaneous breaking of $U(1)_{\rm B-L}$. We discuss implications for the early universe physics when the warm nature of the vector DM serves as a resolution to both the small scale problems that $\Lambda$CDM model encounters and the XENON1T anomaly.
\end{abstract}

\maketitle
\section{Introduction}  
 Recently, the XENON1T collaboration reported an excess in the electronic recoil data for the energy regime ranging from 1\,keV to 7\,keV~\cite{Aprile:2020tmw}. Especially, the prominence of the excess for $m_{A'}\!\in\!(2\,{\rm keV},\!3\,{\rm keV})$ aroused many interesting interpretations based on various extensions of the SM. Absorption of a vector boson is one of the plausible possibilities for the excess in which case its suspected mass and kinetic mixing read $m_{A'}\!\in\!(2\,{\rm keV},\!3\,{\rm keV})$ and $\kappa\!\sim\!10^{-15}$ respectively~\cite{Alonso-Alvarez:2020cdv,Choi:2020udy,An:2020bxd,Nakayama:2020ikz}. Provided this vector boson serves as a dominant component of DM today, its suspected mass regime could be of interest in regard to the small scale problems (e.g. core/cusp problem \cite{Moore:1999gc}, missing satellite problem \cite{Moore:1999nt,Kim:2017iwr}, too-big-to-fail problem \cite{Boylan_Kolchin_2011}); keV scale WDM can alleviate some of the small scale problems if the free-streaming length travelled by the WDM amounts to $\mathcal{O}(0.1)\,{\rm Mpc}$~\cite{Bringmann:2007ft,Cembranos:2005us,Colin:2000dn}. 

Note that, however, the dark photon ($A_{\mu}^{'}$) mass $m_{A'}\!\in\!(2\,{\rm keV},\!3\,{\rm keV})$ is actually outside of the allowed thermal WDM mass regimes inferred from Lyman-$\alpha$ forest observation $m_{\rm wdm}^{\rm thermal}>5.3\,{\rm keV}$~\cite{Irsic:2017ixq} and redshifted 21cm signals in EDGES observations $m_{\rm wdm}^{\rm thermal}>6.1\,{\rm keV}$~\cite{Schneider:2018xba,Lopez-Honorez:2018ipk}. Therefore in order for the dark photon to be a candidate for WDM to address the small scale problems, it should be a non-thermally originated one. Very interestingly, this line of reasoning to have a non-thermal dark photon DM (DPDM) gathers momentum  when it is taken into account that a fit of good quality for the excess can be accomplished for the kinetic mixing  $\kappa\!\sim\!10^{-15}$~\cite{Aprile:2020tmw,Alonso-Alvarez:2020cdv,Choi:2020udy,An:2020bxd}. Being the most direct and dangerous coupling to thermalize $A_{\mu}^{'}$ with the SM thermal bath, the observed kinetic mixing $\kappa\!\sim\!10^{-15}$ ensures that $A_{\mu}^{'}$ can avoid to join the SM thermal bath unless there is another significant indirect coupling to the SM sector. Now there arises an interesting question: what could be a non-thermal production mechanism that allows the mass regime $m_{A'}\!\in\!(2\,{\rm keV},\!3\,{\rm keV})$ consistent with the Lyman-$\alpha$ forest observation? Different non-thermal production mechanisms will be characterized by different momentum spaces for $A_{\mu}^{'}$. And we will invoke the decay of a scalar in the coherent oscillation to answer this question.

In this letter, we study the scenario where the keV-scale DPDM with the suppressed kinetic mixing arises with a non-thermal origin. Especially for $m_{A'}\!\in\!(2\,{\rm keV},\!3\,{\rm keV})$, the DPDM can induce the XENON1T anomaly through absorption analogous to the photoelectric effect. We identify $A_{\mu}^{'}$ with the massive gauge boson of the broken $U(1)_{\rm B-L}$ gauge theory which is the most well-motivated minimal extension of the SM.\footnote{Other examples of the use of $U(1)_{\rm B-L}$ gauge symmetry for addressing the XENON1T anomaly can be found in Refs~\cite{Lindner:2020kko,Gao:2020wfr}.} When incorporated with a scalar and three heavy right handed neutrinos, $U(1)_{\rm B-L}$ gauge theory provides us with the most elegant explanation for the origin of the smallness of the active neutrino masses based on the seesaw mechanism~\cite{Yanagida:1979as,GellMann:1980vs,Minkowski:1977sc} and the leptogenesis~\cite{Fukugita:1986hr,Buchmuller:2005eh}. In spite of this concreteness of the model, our mechanism and result can be easily generalized to a hidden broken $U(1)$ gauge theory, emphasizing its usefulness in the study of the DPDM. To enable the observed mass regime $m_{A'}\!\in\!(2\,{\rm keV},\!3\,{\rm keV})$ to be consistent with the Lyman-$\alpha$ forest observation, we consider the production mechanism where the non-relativistic scalar charged under $U(1)_{\rm B-L}$ produces $A_{\mu}^{'}$ via its decay after $U(1)_{\rm B-L}$ gets spontaneously broken.\footnote{A variety of other ways to produce the vector DM non-thermally has been suggested: resonant decays from an axion-like scalar via a tachyonic instability~\cite{Agrawal:2018vin,Co:2018lka,Bastero-Gil:2018uel,Nakai:2020cfw}, production from a dark Higgs via a parametric resonance~\cite{Dror:2018pdh}, decays from a network of the cosmic strings~\cite{Long:2019lwl}, vector coherent oscillation~\cite{Nelson:2011sf,Arias:2012az,AlonsoAlvarez:2019cgw,Nakayama:2019rhg}, and the gravitational production~\cite{Graham:2015rva,Ema:2019yrd,Ahmed:2020fhc}.} The parent scalar is assumed to obtain a positive Hubble induced mass squared greater than the scalar's mass squared during the inflation and later decays to $A_{\mu}^{'}$ while going through the coherent oscillation. To ensure the consistency with the Lyman-$\alpha$ flux power spectrum, we compute the free-streaming length of $A_{\mu}^{'}$ based on its history given in the model. We show presence of the parameter space in which $A_{\mu}^{'}$ becomes the WDM candidate resolving some of the small scale problems. For defining the WDM, we take $\lambda_{\rm FS}\lesssim1{\rm Mpc}$ as the criterion. Especially $\lambda_{\rm FS}\simeq0.3-0.5{\rm Mpc}$ is of our interest since it is consistent with non-vanishing matter power spectrum at large scales (0.5Mpc) and helps WDM candidate resolve the missing satellite problem (0.3Mpc)~\cite{Bringmann:2007ft,Cembranos:2005us,Colin:2000dn}. 

%%%%%%%%%%%%%%%%%%%%%%%%%%%%%%%%%%%

\section{Model}
\label{sec:model} 
We consider the broken $U(1)_{\rm B-L}$ gauge theory of which the massive gauge boson $A_{\mu}^{'}$ is taken to be the DM candidate in the model. On top of the SM particle contents, we introduce a complex scalar $\Phi$ (-2) and three right-handed neutrino Weyl fields $\overline{N}_{i=1,2,3}$ (+1) where the numbers in the parenthesis denote $U(1)_{\rm B-L}$ charges assigned to each field. These additional fields form the following Yukawa coupling 
\beq
\mathcal{L}_{\rm Yuk}=\sum_{i=1}^{3}\frac{1}{2}y_{N,ij}\Phi\overline{N}^{(i)}\overline{N}^{(j)}+{\rm h.c.}\quad\,,
\label{eq:Nyukawa}
\eeq
by which the right handed neutrinos acquire masses when the condensation of $\Phi$ induces the spontaneous breaking of $U(1)_{\rm B-L}$. Stemming from the large vacuum expectation value (VEV) of $\Phi$, the heaviness of these right handed neutrino fields can explain the tiny active neutrino masses via the seesaw mechanism~\cite{Yanagida:1979as,GellMann:1980vs,Minkowski:1977sc}. Moreover, the out-of-equilibrium decay of the heavy right-handed neutrinos creates the lepton asymmetry of the universe which is converted into the baryon asymmetry later with the help of the sphaleron transition~\cite{Fukugita:1986hr}.

Since $A_{\mu}^{'}$ in the model is identified with the hypothetical vector boson triggering the XENON1T excess, we set $m_{A'}=2g_{\rm B-L}V_{\rm B-L}\simeq2-3{\rm keV}$ where $g_{\rm B-L}$ is the gauge coupling of $U(1)_{\rm B-L}$ and $<\!\!\Phi\!\!>\equiv V_{\rm B-L}/\sqrt{2}$ is defined. As will be shown in the coming sections, the model is to be featured by a high breaking scale $V_{\rm B-L}\gtrsim10^{16}{\rm GeV}$. From $m_{A'}\!\in\!(2\,{\rm keV},\!3\,{\rm keV})$, we obtain $g_{\rm B-L}\lesssim\mathcal{O}(10^{-22})$. Along with these tiny $g_{\rm B-L}$ and $\kappa$, $A_{\mu}^{'}$'s interaction with other particles in the model is extremely feeble and thus $A_{\mu}^{'}$ is totally isolated entity in the model. As the DM candidate, the lifetime of $A_{\mu}^{'}$ is required to be greater than at least the current age of the universe. Indeed, we can see that the requirement $\tau_{A'}>13.8{\rm Gyr}$ is easily satisfied since the above $g_{\rm B-L}$, $V_{\rm B-L}$ and $m_{A'}$ are easily satisfying either of $m_{A'}^{3}<10^{-40}V_{\rm B-L}^{2}{\rm GeV}$ or $g_{\rm B-L}<5\times10^{-18}\times(m_{A'}/1{\rm keV})^{-1/2}$.\footnote{The last inequalities are obtained from $\Gamma(A_{\mu}^{'}\rightarrow f+\bar{f})\lesssim(13.8{\rm Gyr})^{-1}$ where $f$ is a SM fermion with its mass satisfying $m_{A'}\gtrsim2m_{f}$.}

On the other hand, the scalar sector of the model is described by the following potential 
\begin{eqnarray}
V_{\rm scalar}&=&-m_{\Phi}^{2}|\Phi|^{2}+\lambda|\Phi|^{4}\cr\cr
&+&\lambda_{H\Phi}H^{\dagger}H|\Phi|^{2}+V(H)\,,
\label{eq:Vscalar}
\end{eqnarray}
where $H$ denotes the SM SU(2) Higgs doublet and $V(H)$ is the Higgs potential in the SM. We see that there appear two new dimensionless couplings $\lambda$ and $\lambda_{H\Phi}$, and one dimensionful coupling $m_{\Phi}$ as compared to the SM scalar sector. If $\lambda_{H\Phi}$ is significant, our purpose of producing the non-thermal DPDM might be challenged: after $U(1)_{\rm B-L}$ is broken, the longitudinal component ($\theta$) of $A_{\mu}^{'}$ could have couplings to the radial component ($\phi$) of $\Phi$ which is independent of $g_{\rm B-L}$. Thus, once $\phi$ is produced from the SM thermal bath, $\theta$ becomes easily thermalized as well. To avoid this danger, we assume a sufficiently suppressed $\lambda_{H\Phi}$ in Eq.~(\ref{eq:Vscalar}) in this work. On the other hand, differing from $\lambda_{H\Phi}$, there is no any a priori assumption for $m_{\Phi}$ and $\lambda$. In the coming next sections, we study how these new couplings ($\lambda,m_{\Phi}$) are constrained in the light of the purpose to have $A_{\mu}^{'}$ non-thermally produced from $\Phi$ decay and become WDM which can resolve the small scale problems and explain the XENON1T anomaly.

\section{Production of the dark photon}
\label{sec:history}
As a scalar field present during the inflation era, $\Phi$ may obtain the Hubble induced mass much greater than its original mass. For our model, we assume $\Phi$ acquires a positive Hubble induced mass squared  during inflation so that the effective potential of $\Phi$ during inflation becomes
\beqs
\left.V(\Phi)\right\vert_{a<a_{\rm end}}&=&(c_{I}\mathcal{H}_{I}^{2}-m_{\Phi}^{2})|\Phi|^{2}+\lambda|\Phi|^{4}\cr\cr
&\simeq&c_{I}\mathcal{H}_{I}^{2}|\Phi|^{2}+\lambda|\Phi|^{4}\,,
\label{eq:VPhieff}
\eeqs
where $a_{\rm end}$ is the scale factor at which the inflation ends, $c_{I}$ is assumed to be a positive dimensionless coupling and $\mathcal{H}_{I}$ is the Hubble expansion rate during the inflation. $V(\Phi)$ reduces to the second line form due to the assumption of $\mathcal{H}_{I}>\!\!>m_{\Phi}^{2}$. Thus, it can be seen from Eq.~(\ref{eq:VPhieff}) that $\Phi$ sits on the origin in the field space during inflation.

After the inflation ends ($a>a_{\rm end}$), $V(\Phi)$ returns to the form given in the first line of Eq.~(\ref{eq:Vscalar}), i.e. $V(\Phi)=-m_{\Phi}^{2}|\Phi|^{2}+\lambda|\Phi|^{4}$. Then the global minima of the potential becomes located at $|\Phi|=V_{\rm B-L}/\sqrt{2}=m_{\Phi}/\sqrt{2\lambda}$ where $V_{\rm B-L}=m_{\Phi}/\sqrt{\lambda}$. This makes the displacement of $\Phi$ from the global minima at the end of inflation amount to $\Phi_{\rm ini}=V_{\rm B-L}/\sqrt{2}$. Until the time when the Hubble expansion rate ($\mathcal{H}$) becomes comparable to the mass of $\Phi$ is reached, $\Phi$ field stays at the origin in the field space with its energy density $\rho_{\Phi}\simeq(m_{\Phi}^{2}\Phi_{\rm ini}^{2})/2$. From $\mathcal{H}\simeq m_{\Phi}$, we read the temperature of the SM plasma when $\Phi$ field starts its coherent oscillation at $a=a_{\rm osc}$
\beqs
T(a_{\rm osc})&\simeq&\left(\frac{90}{\pi^{2}}\right)^{1/4}g_{*}(a_{\rm osc})^{-1/4}\sqrt{M_{P}m_{\Phi}}\cr\cr&\simeq&(0.85\times10^{9})\left(\frac{g_{*}(a_{\rm osc})}{100}\right)^{-1/4}\left(\frac{m_{\Phi}}{1{\rm GeV}}\right)^{1/2}\,\,{\rm GeV}\,,\nonumber \\
\label{eq:Tosc}
\eeqs
where $M_{P}=2.4\times10^{18}{\rm GeV}$ is the reduced Planck mass and $g_{*}$ is the effective degrees of freedom for the radiation. As the coherent oscillation starts out, $U(1)_{\rm B-L}$ gets spontaneously broken and the heavy right-handed neutrinos acquire masses via the Yukawa coupling given in Eq.~(\ref{eq:Nyukawa}). In this work, we focus on the case where a reheating temperature $T_{\rm RH}$ is greater than $T(a_{\rm osc})$.\footnote{For the other case with $T_{\rm RH}<T(a_{\rm osc})$, we confirmed that presence of a viable parameter space requires very high reheating temperature greater than $10^{17}${\rm GeV}. This may make it difficult to find a consistent inflation model. Thus we restrict ourselves to the case with $T_{\rm RH}>T(a_{\rm osc})$.} 

Since $a=a_{\rm osc}$, the coherent oscillation of $\Phi$ continues in the expanding background and eventually when the Hubble expansion rate becomes comparable to the rate of the decay of $\Phi$ to a pair of longitudinal modes of $A_{\mu}^{'}$, the decay of $\Phi$ produces $A_{\mu}^{'}$.\footnote{For another way of non-thermal DM production mechanism, see, for example, Refs.~\cite{Choi:2020tqp,Choi:2020nan,Choi:2020udy}. } Note that this $A_{\mu}^{'}$ production time coincides with the onset of the free-streaming of $A_{\mu}^{'}$ thanks to the feeble interaction of $A_{\mu}^{'}$ with other particles. By equating the decay rate $\Gamma_{\phi\rightarrow A^{'}\!+\!A^{'}}$ to the Hubble expansion rate during radiation dominated era, we obtain the time ($a=a_{\rm FS}$) of $A_{\mu}^{'}$-production 
\begin{eqnarray}
&&\,\,\Gamma_{\phi\rightarrow A^{'}\!+\!A^{'}}\simeq\frac{Q_{\rm \Phi}^{4}g_{\rm B-L}^{4}V_{\rm B-L}^{2}m_{\phi}^{3}}{32\pi m_{A'}^{4}}\simeq\frac{T^{2}(a_{\rm FS})}{M_{P}}\cr\cr\Longleftrightarrow&&\,\,T(a_{\rm FS})\simeq2.2\times10^{8}\times\sqrt{\lambda}\times\left(\frac{m_{\Phi}}{1{\rm GeV}}\right)^{1/2}{\rm GeV}\,,\nonumber \\
\label{eq:TSMastar}
\end{eqnarray}
where we used $m_{\phi}=\sqrt{2\lambda}V_{\rm B-L}$, $Q_{\Phi}$ is the $U(1)_{\rm B-L}$ charge of $\Phi$, and $a_{\rm FS}$ is the scale factor for the onset of the free-streaming of $A_{\mu}^{'}$.\footnote{For the decay rate formula, we used the mass of the radial component of $\Phi$ which is $m_{\phi}=\sqrt{2}m_{\Phi}$ because it is $|\Phi|$ that decays to the longitudinal mode of $A_{\mu}^{'}$.}  For the second line in Eq.~(\ref{eq:TSMastar}), we used $Q_{\Phi}=-2$. Using the temperature $T_{\rm }(a_{\rm EW})\simeq100{\rm GeV}$ and scale factor $a_{\rm EW}\simeq10^{-15}$ at which the electroweak symmetry breaking takes place, we can estimate $a_{\rm FS}$ 
\beq
a_{\rm FS}\simeq4.5\times10^{-22}\times\lambda^{-1/2}\times\left(\frac{m_{\phi}}{1{\rm GeV}}\right)^{-1/2}\,.
\label{eq:aFS}
\eeq
based on the entropy conservation, i.e. $a_{\rm EW}T(a_{\rm EW})\simeq a_{\rm FS}T(a_{\rm FS})$ and Eq.~(\ref{eq:TSMastar}).

In this section, we studied one possible mechanism of the non-thermal production of $A_{\mu}^{'}$. The feeble interaction between $A_{\mu}^{'}$ and other particles in the model deprives $A_{\mu}^{'}$ of any chance to be produced from the thermal plasma. Nevertheless, the presence of $\Phi$ in the model enables unexpected intriguing way of $A_{\mu}^{'}$-production: the coherently oscillating homogeneous $\Phi$ field decays to produce the longitudinal mode of $A_{\mu}^{'}$.\footnote{The other non-thermal production of $A_{\mu}^{'}$ was considered in Ref.~\cite{Choi:2020dec}. There, the radial component of $\Phi$ is non-thermally produced from the scattering among the right-handed neutrinos via $\overline{N}+\overline{N}\rightarrow\phi+\phi$. These $\phi$ particles either form the dark thermal bath or just free-stream and then $A_{\mu}^{'}$ can be generated from the decay of $\phi$. However, for this scenario, it is figured out that DPDM with mass $2-3$keV cannot be consistent with the Lyman-$\alpha$ forest observation.} In the next section, we constrain the parameter space $(\lambda,m_{\Phi})$ by attributing the current DM population to $A_{\mu}^{'}$ as well as by requiring $A_{\mu}^{'}$ to travel the distance $\lambda_{\rm FS}\lesssim1{\rm Mpc}$ with a special interest in $\lambda_{\rm FS}\simeq0.3-0.5{\rm Mpc}$ since production.

\section{Dark Photon as the WDM}
\label{sec:DPWDM}
Here we study physical quantities that characterize the dark photon $A_{\mu}^{'}$ as the WDM. These quantities include the relic density, $\Delta N_{\rm eff}$ contributed by $A_{\mu}^{'}$ and the free-streaming length $\lambda_{\rm FS}$ of $A_{\mu}^{'}$.
\subsection{Relic density}
\label{sec:DMdensity}
We assume that $A_{\mu}^{'}$ explains the whole of DM population today. The fraction of the energy density of the universe today attributed to the DM reads 
\begin{equation}
\Omega_{\rm DM,0}=\frac{\rho_{\rm DM,0}}{\rho_{\rm cr,0}}=\frac{s_{0}}{\rho_{\rm cr,0}}\times m_{\rm DM}\times Y_{\rm DM}\simeq0.27\,,
\label{eq:DMdensity}
\end{equation}
where the comoving number density of DM, $Y_{\rm DM}\equiv n_{\rm DM}/s$, is the conserved quantity since the production of the DM. From the values of $s_{0}=2.21\times10^{-11}{\rm eV}^{3}$ and $\rho_{\rm cr,0}=8.02764\times10^{-11}\times h^{2}\,{\rm eV}^{4}$, one obtains
\begin{equation}
    Y_{\rm DM}=9.8\times10^{-4}\times\left(\frac{m_{\rm DM}}{1{\rm keV}}\right)^{-1}\times h^{-2}\,,
\label{eq:YDM1}
\end{equation}
where $h$ is defined to be $\mathcal{H}_{0}=100h{\rm km/sec/Mpc}$. On the other hand, having the decay process $\phi\!\rightarrow\! A_{\mu}^{'}+A_{\mu}^{'}$ as the origin of DM in the model, we have the relation $n_{A'}=2n_{\phi}(=2n_{\Phi})$. $\Phi$'s number density can be estimated by $n_{\Phi}=\rho_{\Phi}/m_{\Phi}$. Using these relations, for the case where $T_{\rm RH}>T(a_{\rm osc})$, we can estimate the DPDM comoving number density as what follows
\beqs
Y_{\rm DM}&\equiv&\frac{n_{A'}}{s}=\left.\frac{2n_{\Phi}}{s}\right\vert_{a=a_{\rm osc}}\cr\cr
&=&\frac{m_{\Phi}\Phi_{\rm ini}^{2}}{\frac{2\pi^{2}}{45}g_{*,s}(a_{\rm osc})T(a_{\rm osc})^{3}}\,,
\label{eq:Yphi}
\eeqs
where $s$ is the entropy density of the universe and $\Phi_{\rm ini}=V_{\rm B-L}/\sqrt{2}$ is assumed. We used the fact that the comoving number density of $\phi$ remains conserved since $\phi$ field begins the oscillation. Thus, using $h=0.68$, equating Eq.~(\ref{eq:YDM1}) and Eq.~(\ref{eq:Yphi}) yield
\beq
\left(\frac{g_{*}(a_{\rm osc})}{100}\right)^{1/4}\!\!\!\left(\frac{m_{\Phi}}{1{\rm GeV}}\right)^{1/2}\!\!\!\left(\frac{m_{A'}}{1{\rm keV}}\right)^{-1}\!\!\!\left(\frac{V_{\rm B-L}}{10^{13}{\rm GeV}}\right)^{-2}\!\!\simeq0.875\,,
\label{eq:Yphi2}
\eeq
where we used $g_{*}(a_{\rm osc})=g_{*,s}(a_{\rm osc})$. In Fig.~\ref{fig1}, the set of points on $(\lambda,m_{\Phi})$ plane obeying Eq.~(\ref{eq:Yphi2}) is shown as the red solid line.

\subsection{$\Delta N_{\rm eff}$ contributed by the dark photon}
\begin{figure}[t]
\centering
\hspace*{-5mm}
\includegraphics[width=0.45\textwidth]{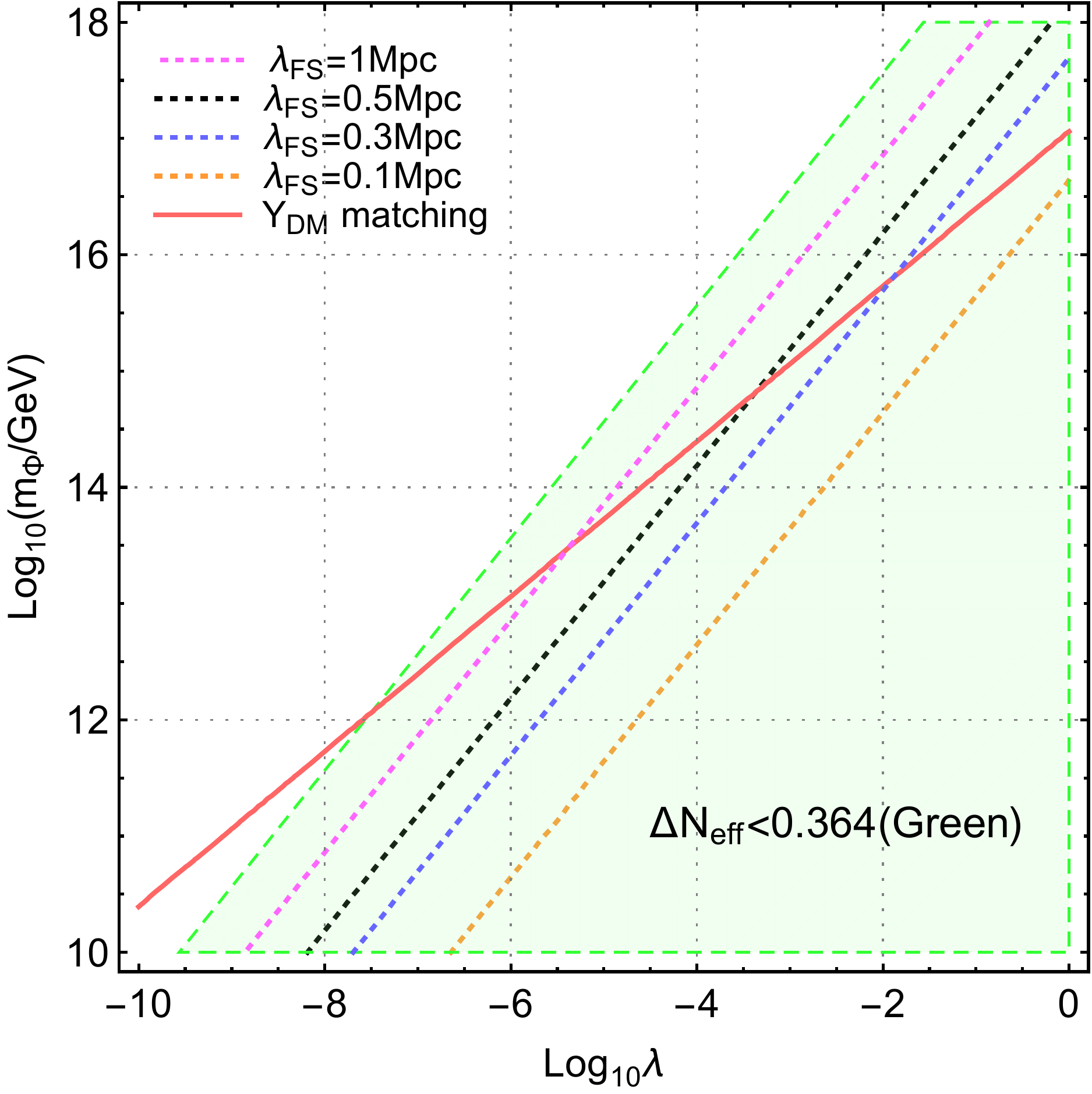}
\caption{The red solid line is the set of points on $(\lambda,m_{\Phi})$ plane producing $A_{\mu}^{'}$ which can explain the current DM relic density. Each dotted line of different color is the set of points ($\lambda,m_{\Phi}$) making DPDM travel the specified free-streaming length. For the points within the green shaded region, DPDM's contribution to the radiation energy density at BBN era satisfies $\Delta N_{\rm eff}^{\rm BBN}<0.364$. Here the relation $m_{\phi}=\sqrt{2}m_{\Phi}$ for converting the constraint on $m_{\phi}$ to that on $m_{\Phi}$ is used.}
\vspace*{-1.5mm}
\label{fig1}
\end{figure}

As the keV-scale DM candidate, $A_{\mu}^{'}$ is the relativistic particle during the BBN era, behaving as the radiation. Therefore, its energy density contributes to the background expansion during the radiation dominated era, which is parametrized by its contribution to the extra effective number of neutrinos, $\Delta N_{\rm eff}^{\rm BBN}$. The momentum of $A_{\mu}^{'}$ when produced is well defined to be $p_{A'}(a_{\rm FS})=m_{\phi}/2$ because the mother scalar field is homogeneous and behaves like a motionless matter. The amount of energy density of $A_{\mu}^{'}$ at BBN time is given by
\begin{eqnarray}
\rho_{{\rm DM}}(a_{\rm BBN})&=&\sqrt{m_{{\rm DM}}^{2}+\left(\frac{m_{\phi}a_{\rm FS}}{2a_{\rm BBN}}\right)^{2}}\times Y_{\rm DM}\cr\cr
&&\times\frac{2\pi^{2}}{45}g_{s}(a_{{\rm BBN}})T(a_{\rm BBN})^{3}\,,
\label{eq:rhoDM}
\end{eqnarray}
where $Y_{\rm DM}$ is given in Eq.~(\ref{eq:YDM1}) and $g_{s}$ is the effective number of relativistic degrees of freedom for the entropy density. In terms of $\rho_{\rm DM}(a_{\rm BBN})$, $\Delta N_{\rm eff}^{\rm BBN}$ is defined to be
\beq
\Delta N_{\rm eff}^{\rm BBN}\simeq\frac{\rho_{{\rm DM}}(a_{\rm BBN})}{\rho_{\gamma}(a_{\rm BBN})}\times\frac{8}{7}\left(\frac{11}{4}\right)^{4/3}\,.
\label{eq:Neff}
\eeq
From the constraint on $\Delta N_{\rm fluid}^{\rm BBN}\!\leq\!0.364$~\cite{Cyburt:2015mya}(95\% C.L.), Eq.~(\ref{eq:rhoDM}) and Eq.~(\ref{eq:Neff}), we obtain
\beq
m_{\phi}\lesssim(2.3-5.2)\times10^{19}\lambda\,\,{\rm GeV}\,,
\label{eq:mphiaFS}
\eeq
for $m_{A'}\!\in\!(2\,{\rm keV},\!3\,{\rm keV})$. The green shaded region in Fig.~\ref{fig1} satisfies Eq.~(\ref{eq:mphiaFS}).

\subsection{Free-streaming length of the dark photon}
As the WDM candidate, $A_{\mu}^{'}$ can alleviate the small scale problems provided its free-streaming length amounts to $\lambda_{\rm FS}\simeq\mathcal{O}(0.1){\rm Mpc}$. As mentioned in the introduction, we focus on the parameter space which can produce $\lambda_{\rm FS}\lesssim1{\rm Mpc}$ with a special interest in $\lambda_{\rm FS}=0.3-0.5{\rm Mpc}$ to make $A_{\mu}^{'}$ DPDM WDM candidate addressing the small scale problems, still consistent with the non-vanishing matter power spectrum at the large scales~\cite{Bringmann:2007ft,Cembranos:2005us,Colin:2000dn}. 

The free-streaming length is computed by
\begin{eqnarray}
\lambda_{\rm FS}&=&\int_{t_{\rm FS}}^{t_{0}}\frac{<\!\!v_{A'}(t)\!\!>}{a}{\rm d}t\cr\cr
&=&\int_{a_{\rm FS}}^{1}\frac{{\rm d}a}{\mathcal{H}_{0}F(a)}\frac{<\!\!p_{A'}(a_{\rm FS})\!\!>a_{\rm FS}}{\sqrt{(<\!\!p_{A'}(a_{\rm FS})\!\!>a_{\rm FS})^{2}+m_{\rm DM}^{2}a^{2}}}\,, \nonumber \\
\label{eq:FSL}
\end{eqnarray}
where $F(a)\equiv\sqrt{\Omega_{\rm rad,0}+a\Omega_{\rm m,0}+a^{4}\Omega_{\Lambda,0}}$, $a_{\rm FS}$ is given in Eq.~(\ref{eq:aFS}) and $<\!\!p_{A'}(a_{\rm FS})\!\!>\simeq m_{\phi}/2$. Importantly, for the case where $A'$ is produced from the coherently oscillating scalar's decay, the momentum space of $A'$ is expected to be almost close to the delta function centered around $m_{\phi}/2$ since the mother scalar field has almost zero momentum. This makes the estimation of $\lambda_{\rm FS}$ of $A_{\mu}^{'}$ in our model much more reliable than that for $A_{\mu}^{'}$ with a significant width of the momentum space distribution. Because $a_{\rm FS}$ is a function of $(\lambda,m_{\Phi})$, so is $\lambda_{\rm FS}$ in our model. In Fig.~\ref{fig1}, we show the set of points producing $\lambda_{\rm FS}=0.1,0.3,0.5$ and 1Mpc as the dotted yellow, blue, black and magenta lines respectively.

\subsection{Combined constraints on $V(\Phi)$}
In Fig.~\ref{fig1}, for the exemplary value of $m_{A'}=3{\rm keV}$, we show the resultant parameter space of $(m_{\Phi},\lambda)$ obtained by applying the constraints on $\Delta N_{\rm eff}^{\rm BBN}$ and the free-streaming criterion. The green shaded area shows the region where $\Delta N_{\rm eff}^{\rm BBN}\lesssim0.364$ is satisfied. Each of dotted lines of different colors is a set of points on $(m_{\Phi},\lambda)$ plane producing each specified free-streaming length of $A_{\mu}^{'}$ based on Eq.~(\ref{eq:FSL}). The red line shows the set of points which makes the abundance of $A_{\mu}^{'}$ explain the current DM relic density.

It can be seen from Fig.~\ref{fig1} that those DPDMs arising from the decay of the mother scalar with the mass $m_{\Phi}\gtrsim10^{13}{\rm GeV}$ and the quartic coupling $\lambda\gtrsim5\times10^{-6}$ can not only explain the current DM relic density but also resolve the small scale problems as the WDM by travelling the distance $\lambda_{\rm FS}\lesssim1{\rm Mpc}$ since production. As such, $A_{\mu}^{'}$ in the model is really shown to be a non-thermally originated WDM causing XENON1T anomaly, consistent with the various existing cosmological constraints.

From $T_{\rm RH}>T(a_{\rm osc})\simeq\sqrt{m_{\Phi M_{P}}}$ and the viable scalar mass regime $m_{\Phi}\gtrsim10^{13}{\rm GeV}$, we can infer that the model is consistent with a reheating temperature as high as $T_{\rm RH}\simeq10^{15}{\rm GeV}$. If one focuses on the regime $\lambda_{\rm FS}\simeq0.3-0.5{\rm Mpc}$, then consistent $T_{\rm RH}$ becomes required to be as high as $T_{\rm RH}\simeq10^{16}{\rm GeV}$.

Finally, the obtained viable range of $m_{\Phi}$ and $\lambda$ indicates the breaking scale $V_{\rm B-L}\simeq\mathcal{O}(10^{16}){\rm GeV}$. Given this scale, we found that the right-handed neutrino mass satisfying $m_{\bar{N}}\lesssim10^{12}{\rm GeV}$ can be consistent with our DM production scenario without thermalizing $\phi$ particle. This result, combined with the $T_{\rm RH}$ constraint above, indicates that the leptogenesis is realized in the model in the thermal manner (thermal leptogenesis).

\section{Discussion}
In this letter, motivated by the recently reported XENON1T excess, we propose a minimal model where the massive gauge boson of $U(1)_{\rm B-L}$ gauge theory with the kinetic mixing $\simeq10^{-15}$ can play the role of the dark photon dark matter with its mass $m_{A'}\!\in\!(2\,{\rm keV},\!3\,{\rm keV})$ inducing XENON1T anomaly. As the DM candidate with the free-streaming length of order $\sim\mathcal{O}(0.1){\rm Mpc}$, $A_{\mu}^{'}$ is classified as the WDM which can address the small scale problems that $\Lambda$CDM is suffering from. 

Relying on $U(1)_{\rm B-L}$ gauge theory which is the well motivated extension of the SM, we introduced a scalar and three heavy right-handed neutrinos charged under $U(1)_{\rm B-L}$ in addition to the SM particle content, which is the minimal set-up for ensuring the successful operation of the seesaw mechanism and leptogenesis. Imposing masses to the three-right handed neutrinos, the scalar was shown to be able to serve as the parent particle for the dark photon warm dark matter. Due to the non-thermal production mechanism we assumed in the model, the gauge boson mass of $m_{A'}\in(2{\rm keV},3{\rm keV})$ can be consistent with the Lyman-$\alpha$ forest observation. Intriguingly, we figured out that for the self-quartic interaction $\lambda\gtrsim5\times10^{-6}$ and the scalar mass $m_{\Phi}\gtrsim10^{13}{\rm GeV}$, the model can  produce the non-thermally originated dark photon warm dark matter without running a foul of existing cosmological and astrophysical constraints as a source of XENON1T anomaly. Due to the inferred heavy mass of $\Phi$, the model is found to demand a high scale inflation with $\mathcal{H}_{I}\gtrsim\mathcal{O}(10^{13}){\rm GeV}$ for consistency. Although we focus on $U(1)_{\rm B-L}$ gauge theory, our DPDM production mechanism based on the decay of the scalar going through the coherent oscillation can be easily applied to a hidden Abelian gauge theory. \\

{\bf Note added}: After we finished this paper, we became aware of arXiv:2007.02898 [hep-ph]~\cite{Okada:2020evk}. However, the mechanism to make the $U(1)_{\rm B-L}$ gauge boson DM there is different from ours. See also references in their paper for proposals related to XENON1T anomaly.

% ==================================================================

\begin{acknowledgments}
This work is dedicated to Roberto Peccei. N. Y. is supported by JSPS KAKENHI Grant Number JP16H06492.
T. T. Y. is supported in part by the China Grant for Talent Scientific Start-Up Project and the JSPS Grant-in-Aid for Scientific Research No. 16H02176, No. 17H02878, and No. 19H05810 and by World Premier International Research Center Initiative (WPI Initiative), MEXT, Japan. 

\end{acknowledgments}

% ================================================================

\bibliography{main}

\end{document}